# Transferring Orbital Angular Momentum to an Electron Beam Reveals Toroidal and Chiral Order

## Authors

K. X. Nguyen[1*], Y. Jiang[2], M. C. Cao[3], P. Purohit[2], A. K. Yadav[4], P. García-Fernández[5], M. W. Tate[2], C. S. Chang[2], P. Aguado-Puente[6], J. Íñiguez[7,8], F. Gomez-Ortiz[5], S. M. Gruner[2,9,10], J. Junquera[5], L. W. Martin[11,12], R. Ramesh[11,13], D. A. Muller[3,9*]

*To whom correspondence should be addressed:
Email: kn324@cornell.edu, david.a.muller@cornell.edu

## Affiliations

[1.] Department of Chemistry and Chemical Biology, Cornell University, Ithaca, NY, USA
[2.] Department of Physics, Cornell University, Ithaca, NY, USA
[3.] School of Applied and Engineering Physics, Cornell University, Ithaca, NY, USA
[4.] Department of Electrical Engineering and Computer Sciences, University of California
[5.] Departamento de Ciencias de la Tierra y Física de la Materia Condensada, Universidad de Cantabria, Cantabria Campus Internacional, Avenida de los Castros s/n, 39005, Santander, Spain
[6.] Atomistic Simulation Center, Queen's University Belfast, Belfast BT7 1NN, United Kingdom
[7.] Materials Research and Technology Department, Luxembourg Institute of Science and Technology (LIST), 5 avenue des Hauts-Fourneaux, L-4362 Esch/Alzette, Luxembourg
[8.] Physics and Materials Science Research Unit, University of Luxembourg, 41 Rue du Brill, L-4422 Belvaux, Luxembourg
[9.] Kavli Institute at Cornell for Nanoscale Science, Ithaca, NY USA
[10.] Cornell High Energy Synchrotron Source (CHESS), Cornell University, Ithaca, NY, USA
[11.] Department of Material Science and Engineering, University of California, Berkeley, Berkeley, California 94720 USA
[12.] Materials Sciences Division, Lawrence Berkeley National Laboratory, Berkeley, California 94720, USA
[13.] Department of Physics, University of California, Berkeley, California 94270, USA

## Abstract

**Orbital angular momentum and torque transfer play central roles in a wide range of magnetic textures and devices including skyrmions and spin-torque electronics(*1-4*). Analogous topological structures are now also being explored in ferroelectrics, including polarization vortex arrays in ferroelectric/dielectric superlattices(*5*). Unlike magnetic toroidal order, electric toroidal order does not couple directly to linear external fields. To develop a mechanism that can control switching in polarization vortices, we utilize a high-energy electron beam and show that transverse currents are generated by polar order in the ballistic limit. We find that the presence of an electric toroidal moment in a ferro-rotational phase transfers a measurable torque and orbital angular momentum to the electron beam. Furthermore, we find that the complex polarization patterns, observed in these heterostructures, are microscopically chiral with a non-trivial axial component of the polarization. This chirality opens the door for the coupling of ferroelectric and optical properties.**

# Introduction

Topological states of electrical polarization have emerged as an area of research having both fundamental(*6*) and technological relevance(*7*). While absent in bulk, non-trivial polarization field textures, including structures with ferroaxial or toroidal order parameters, were predicted to arise from the close interplay between the geometry of low-dimensional structures and the topology of the polarization field $\boldsymbol{P}(\boldsymbol{r})$(*8, 9*), where $\boldsymbol{r}$ locates the position of the local polarization. The resulting vortex-like topological features which arise from continuous rotations of the polarization field can be characterized by the electric toroidal moment order parameter, $\boldsymbol{g} = \frac{1}{2}\int \boldsymbol{r} \times \boldsymbol{P}(\boldsymbol{r}) d^3\boldsymbol{r}$(*10-12*), giving rise to a host of new coupling mechanisms (*e.g.*, pyrotoroidic, piezotoroidic, electric toroidal susceptibility). Theoretical studies have suggested possible routes to control ferroelectric toroidal order parameters in ferroelectric nanostructures(*13-17*), and recently ordered, vortex-like textures of electrical polarization have been experimentally stabilized(*5*). These were achieved in $(SrTiO_3)_x/(PbTiO_3)_x$ superlattices(*5*), where the balance between electrostatic and strain boundary conditions resulted in nanometer-scale polarization vortex arrays possessing an electric toroidal moment. Soft X-ray dichroism measured on these samples has been interpreted as resulting from an underlying macroscopic chirality from ordered regions of vortex-like polar line defects(*18*). While these studies provide a spatially averaged measurement of the ferrotoroidal state, the microscopic details of the chiral nature of this phase are still unclear.

Here, we show how the toroidal moment can be directly measured quantitatively, with high sensitivity and spatial resolution, using (i) a new generation of momentum-resolved electron microscope pixel array detectors, and (ii) the proportional relationship between the orbital angular momentum (OAM) transferred to the electron beam, $\boldsymbol{L}$, and the toroidal moment of the superlattice, $\boldsymbol{g}$. Our technique can recover torque transfer and OAM with high fidelity and resolution, without compromising the beam shape, and in a geometry where it is possible to simultaneously measure and distinguish electric fields, polarity, and crystal tilt, the latter being a serious challenge for traditional approaches such as holography and differential phase contrast(*19*).

# Results

To date, the electron microscopy approach to measuring OAM has been to start with a beam that has been structured with special apertures to possess a well-defined OAM, such as a vortex beam. In this approach, the vortex beam scatters through the sample and a change is recorded with a localized detector. This approach has been used for magnetic measurements from inelastic scattering(*20-23*), but precludes the simultaneous momentum measurements needed to reliably recover polarity in ferroelectrics. Our approach is, in a sense, to run this experiment backwards, *i.e.*, using a simple and local beam (*i.e.*, with zero OAM), and having it scatter through a sample that has vorticity where the final OAM is recorded with an angle-resolved and phase-sensitive detection method. This scheme is provided by a four-dimensional scanning transmission electron microscopy (4D-STEM) with the electron microscope pixel array detector (EMPAD)(*24*), schematized for the case of a $(PbTiO_3)_{12}/(SrTiO_3)_{12}$ superlattice in Fig. 1A. In this way, we can simultaneously measure the linear momentum transfer to the scattered electron beam $\langle\boldsymbol{p}\rangle$, which recovers polarity(*25, 26*) by measuring the probability current flow(*27-29*) in ferroelectric polar vortices (Fig. 1B-D), and develop an approach to measure angular momentum, detailed in the following section. The polarization texture, where the local dipoles within the $PbTiO_3$ layer continuously rotate forming a sequence of clock-wise/counter-clockwise array of vortices along the [100] direction, obtained with this method is perfectly compatible with the one observed by high-resolution transmission electron microscopy(*5*).

Utilizing information from 4D-STEM, we use ptychography(*30-32*) to first calculate OAM by explicitly calculating

$$\langle L \rangle = \langle \Psi | \hat{r} \times \hat{p} | \Psi \rangle = \int \Psi^*(r)[\hat{r} \times \hat{p}]\Psi(r)dr \qquad (1),$$

where $\Psi$ is the wave function of the electron beam, $r$, is the probe position in real space, and $\hat{r}$ and $\hat{p}$, are the position and momentum operators respectively. Experimentally, within non-overlapping diffraction disks, there is a finite width to their size and an intrinsic uncertainty, such that a relation between $\hat{r}$ and $\hat{p}$ becomes measurable. Furthermore, we find that this method recovers the phase and amplitude of the exit wave function and is exact for all sample thicknesses to within the accuracy of the ptychographic reconstruction. However, ptychography is computationally intensive and requires high sampling densities in real space.

Here, we also propose a faster and more efficient two-step approach that is less restricted by in-plane sampling requirements. The first step is a change in the focus: from the direct measurement of the OAM, to the change (derivative) of its expectation value. This is given by the torque exercised by the sample, $\langle \Gamma \rangle$, given by,

$$\langle \Gamma \rangle = \frac{d\langle L \rangle}{dt} = \langle \Psi | [r \times (-\nabla V)] | \Psi \rangle \qquad (2),$$

where $V$ is the sample potential (see Methods, Eq. S4)(33).

The second step is to take the integral over time of $\langle \Gamma \rangle$ in Eq. (2) to get the total change of the expectation value of $\langle L \rangle$. For elastic scattering, the electron travels at constant velocity where the integration over time for the propagation of the wave packet through the sample becomes an integration over sample thickness. We then use the strong phase approximation to connect the probability current images $\langle p \rangle$ (Fig. 2B) to $\nabla V$ (see supplementary materials, Eq. S8-S18)(34-38). The result provides the z-component of the torque (in the microscope's coordinate system) convolved incoherently with the probe shape, where the electron trajectory as a diffraction-limited electron probe with incident wavepacket, $\Psi_0(\vec{r} - \vec{r}_p)$ centered about incident probe position, $\vec{r}_p$:

$$\langle \Gamma_z(r_p) \rangle = \int \left[(\vec{r} - \vec{r}_p) \times (-\vec{\nabla} V(\vec{r}))\right]_z |\Psi_0(\vec{r} - \vec{r}_p)|^2 d\vec{r} \qquad (3)$$

which can be separated in Fourier space, and then corrected for probe shape (equation S18). Since we image in projection, the beam's passage through the sample gives $\langle L_z \rangle = \int \langle \Gamma_z \rangle dt$.

To test the numerical accuracy of both ptychographic and torque transfer approaches, we perform the multislice simulations(39) on the model polarization vortex structures. Fig. 2C shows the change in OAM calculated using ptychography and Fig. 2D is the total torque transfer, $\langle L_z \rangle$, calculated from $\langle p_x \rangle$ and $\langle p_y \rangle$ images using equation S18; here, we find that there is good agreement between the angular momentum and torque transfer approaches for a moderate thickness sample (< 20 nm). In thick simulated samples, the two approaches begin to diverge as the strong phase approximation breaks down once beam propagation effects become significant (Fig. S1).

We then utilize our faster, efficient torque transfer approach on experimental results of $(PbTiO_3)_{12}/(SrTiO_3)_{12}$ superlattice. First, we plotted the {200} probability current in Fig. 3A as a vector map to show the ordered arrays of polarization vortices, where we find the vortices have offset cores and slightly asymmetric shapes. Second, we show the measured $\langle L_z \rangle$ (Fig. 3B) using the total torque transfer (equation S18) from the same region as Fig. 3A at high magnification and low magnification in Fig. 3C, where we make picometer-precision measurements over arbitrarily large fields of view. We find that our detection scheme overcomes a limitation of real-space imaging; here, the sensitivity is set by the SNR on the detector in momentum space, rather than the picometer-scale instabilities in the scan position–of the electron beam, since it is no longer necessary to resolve and count individual atoms. Furthermore, our approach measures both linear and angular momentum, where $\langle \boldsymbol{L} \rangle$ takes on an additional significance, as it is proportional to the toroidal moment and order parameter $\boldsymbol{g}$ (see Materials and Methods) for ferroaxial textures. As a consequence, we find that we can utilize this approach to uncover the theoretically predicted(40) chiral nature of the vortex states. Here, Fig. 3B shows the toroidal ordering is that of off-centered, alternating, and asymmetric vortices that lack an axis of symmetry. This is a necessary, but not sufficient, condition for chirality.

To determine if the vortex structures are chiral, we need to also investigate if there is a net polarization along the axial direction of the vortex. To do this, we prepared plan-view thinned samples of the $(PbTiO_3)_{12}/(SrTiO_3)_{12}$ superlattice and imaged down the [001] zone of the superlattice (Fig. 4A). In this orientation, the net polarity needed for a chiral structure will appear as a non-zero $\langle p_y \rangle$ component on the EMPAD (again $x$ and $y$ are in the detector coordinate system with $y$ along the axial direction of the vortices). Fig. 4B and 4C show $\langle p_x \rangle$ and $\langle p_y \rangle$ images, respectively, where a small, but non-zero $\langle p_y \rangle$ is indeed detected, having 6 times less intensity than $\langle p_x \rangle$ (Fig. S2). The reduced axial intensity is a consequence of the strong dechanneling of the electron beam on lead atom columns (Fig. S3) which means we do not sample all depths through the sample with equal weighting; here, we observed that most of our signal for $\langle p_x \rangle$ and $\langle p_y \rangle$ comes from the top half of the vortices (shown as the red-shaded region in Fig. 4A). This reduces sensitivity to the axial component and is present both for our method and the less-sensitive high-angle annular dark field (ADF)-based method to measure polar displacements. By comparing second-principles simulations of the projected polarization for left-handed and right-handed chiral structures (Figs. 4D and 4E) we again see the polarization weighted from beam propagation towards the entrance surface. Here, we observed that a left-handed chiral structure show $\langle p_x \rangle$ and $\langle p_y \rangle$ components out of phase whereas a right-handed chiral structure has them tracking one another. Experimental results for $\langle p_x \rangle$ and $\langle p_y \rangle$ (Figs. 4B and 4C) are superimposed to visualize their relative alignments in (Fig. 4H). We observed both left- and right-handed chiral domains (Fig. 4F and 4G, respectively); the small domain size (50-100 nm) could account for the weak chiral signal observed by X-ray scattering[17] where the superposition of both domains would reduce the net macroscopic chiral signal. In contrast, from both our simulation and experiment, we find that this should be a strongly chiral material within each domain, but overall exhibiting both right-handed and left-handed chiral domains in close proximity.

## Discussion

By applying our new detection methods for mapping polar and toroidal order, we directly observe the emergence of chirality in such vortices. This is a direct experimental demonstration that one can take two materials that, by themselves are non-handed, but when assembled under certain boundary conditions in which the two primary energy scales (the elastic and electrostatic energies) are almost of the same order of magnitude, they compete with one another. This leads to order parameter topologies that are chiral with a characteristic length scale of 5-10 nm. If the chirality can be controlled with an electric field or strain, this would point to pathways by which the chirality can be used as an independent order parameter. Furthermore, it is quite possible that the chiral order either co-exists with or emerges from a polar order, thus presenting the possibility for such an electrically controllable phase transition giving rise to a host of coupling phenomena between the toroidal order and electric and mechanical degrees of freedom.

Finally, we note that although the calculations of torque transfer were performed for high-energy electrons, the same symmetry elements and invariants are also present in Bloch-wave theory(41), suggesting that a similar scattering mechanism may be detectable with low-energy (i.e. few eV) electrons as well, giving a new electrical read-out mechanism for toroidal order that would be useful for interrogating a ferroelectric equivalent of a magnetic racetrack memory.

## Materials and Methods

**Scanning Transmission Electron Microscopy (STEM)**
Superlattice samples were imaged using both a 200 keV uncorrected FEI Tecnai F20 and a 60-300 keV probe-corrected FEI Titan Themis. The same EMPAD detector was used on both instruments. Atomic resolution imaging was performed on the Titan instrument at 300 keV with a 29 mrad convergence angle.

For reasonably thin specimens where the strong phase approximation holds (equation S7), the probability current images can be further simplified as a convolution of the incident beam shape and the gradient of the sample potential $V(\vec{r}_p)$(*42*):

$$\langle \vec{p}(\vec{r}_p) \rangle = \hbar\sigma |\Psi_0(\vec{r}_p)|^2 \otimes \vec{\nabla} V(\vec{r}_p) \qquad (4).$$

From which we can now calculate, and measure the OAM, $\langle L \rangle$. We note that $\langle L \rangle (= \hat{r} \times \hat{p})$ is proportional to the toroidal moment and order parameter $g = \frac{1}{2}\int r \times P(r) d^3r$, where $P(r)$ is local dipole density(*10*), as $P(r)$ is proportional to the $\langle p \rangle$ constructed from Friedel pairs in thin samples. This can be seen numerically in Figure 2a vs b where the polarization field tracks the contrast in the probability current images $\langle p_x \rangle$ and $\langle p_z \rangle$.

Electron channeling plays an important role in the depth dependence of the probability current signal where the signal is not a simple projection through the sample. From multislice simulations, the signal from the strongly-scattering Pb columns comes mostly from the top half of the vortex (Extended Data Fig. 3) and is scattered away after that. The dechanneling in the two in-plane directions is quite similar as it is dominated by the atomic column more so than the small displacements, where so much of the thickness variation can be compensated by comparing the relative intensities of the $\langle p_x \rangle$ and $\langle p_y \rangle$ components (Fig. S1).

**PbTiO$_3$/SrTiO$_3$ Growth**
Pulsed Laser Deposition (PLD) was used to synthesize superlattice films of PbTiO$_3$/SrTiO$_3$. All films were grown on SrRuO$_3$-buffered (110) oriented DyScO$_3$ single crystalline substrate. Reflection High Energy Electron Diffraction (RHEED) was used to monitor the growth dynamics of PbTiO$_3$ and SrTiO$_3$. The growth conditions were carefully optimized to obtain layer-by-layer (Frank-van der Merwe) growth of PbTiO$_3$ and SrTiO$_3$, which was sustained for the entire growth of 100 nm thick superlattice film. For detailed account on growth conditions and optimization of other parameters, see Methods section of reference (*5*).

**Second principles simulations of PbTiO$_3$/SrTiO$_3$ superlattices**
The superlattice potentials are identical to those used in Ref. (*43*). The interactions within the PbTiO$_3$ or SrTiO$_3$ layers were based on the previously introduced potentials for the bulk compounds, which give a qualitatively correct description of the lattice dynamical properties and structural phase transitions of both materials. For the interactions between ion pairs at the interface simple numerical averages were used. For the periodicities of the superlattices studied in this work, the main effects of the stacking are purely electrostatic. Those long-range dipole-dipole interactions are governed by the Born effective charges of the bulk parent compounds and a bare electronic dielectric constant $\varepsilon^\infty$ that is taken as a weighted average of the first-principles results for bulk PbTiO$_3$ ($\varepsilon^{\infty,\,PTO}$ = 8.5) and SrTiO$_3$ ($\varepsilon^{\infty,\,STO}$ = 6.2) with weights reflecting the composition of the superlattice. In order to preserve the electrostatic interactions within each material as close as possible to the bulk parent compounds, we have rescaled the Born effective charge tensors of the inner atoms by $\sqrt{\varepsilon^\infty/\varepsilon^{\infty,ABO_3}}$ (where $ABO_3$ stands for PbTiO$_3$ or SrTiO$_3$ depending on the layer to which the atom belongs). In this way, following Eq. (23) of Ref. (*44*), the dipole-dipole interactions remain the same as in bulk even if we adopt a common value of $\varepsilon^\infty$ for the whole heterostructure. The Born tensors corresponding to the atoms at the interfaces were left untouched.

We assume in-plane lattice constants of $a = b = 3.901$ Å and $\gamma = 90°$. To counteract the underestimate of the lattice constant due to the well-known overbinding error of the local density approximation, which is the first-principles theory used to compute the parameters of our model, an external expansive hydrostatic pressure of -11.2 GPa is imposed. These approximations and adjustments allow us to construct models for superlattices of arbitrary $n$ stacking. For computational feasibility, we have focused on a simulation supercell made from a periodic repetition of $2n \times n \times 2n$ elemental perovskite unit cells, sufficiently large to simulate domains in the $n = 10$ superlattice. We solved the models by running Monte Carlo simulations typically comprising 10,000 thermalization sweeps followed by 40,000 sweeps to compute thermal averages. We ran Monte Carlo simulated annealing down to very low temperatures to perform structural relaxations and find the ground state or metastable solutions.

# References

1. L. Berger, Emission of spin waves by a magnetic multilayer traversed by a current. *Physical Review B* **54**, 9353-9358 (1996).
2. E. B. Myers, D. C. Ralph, J. A. Katine, R. N. Louie, R. A. Buhrman, Current-induced switching of domains in magnetic multilayer devices. *Science* **285**, 867-870 (1999).
3. J. C. Slonczewski, Current-driven excitation of magnetic multilayers. *Journal of Magnetism and Magnetic Materials* **159**, L1-L7 (1996).
4. M. Tsoi *et al.*, Excitation of a Magnetic Multilayer by an Electric Current. *Physical Review Letters* **80**, 4281-4284 (1998).
5. A. K. Yadav *et al.*, Observation of polar vortices in oxide superlattices. *Nature* **530**, 198-201 (2016).
6. N. Choudhury, L. Walizer, S. Lisenkov, L. Bellaiche, Geometric frustration in compositionally modulated ferroelectrics. *Nature* **470**, 513-517 (2011).
7. L. W. Martin, A. M. Rappe, Thin-film ferroelectric materials and their applications. *Nature Reviews Materials* **2**, (2016).
8. Y. Nahas *et al.*, Discovery of stable skyrmionic state in ferroelectric nanocomposites. *Nat Commun* **6**, (2015).
9. J. Hong, G. Catalan, D. N. Fang, E. Artacho, J. F. Scott, Topology of the polarization field in ferroelectric nanowires from first principles. *Physical Review B* **81**, 172101 (2010).
10. S. Prosandeev, I. Ponomareva, I. Naumov, I. Kornev, L. Bellaiche, Original properties of dipole vortices in zero-dimensional ferroelectrics. *Journal of Physics: Condensed Matter* **20**, 193201 (2008).
11. S. Hans, Some symmetry aspects of ferroics and single phase multiferroics *. *Journal of Physics: Condensed Matter* **20**, 434201 (2008).
12. I. I. Naumov, L. Bellaiche, H. Fu, Unusual phase transitions in ferroelectric nanodisks and nanorods. *Nature* **432**, 737-740 (2004).
13. L. Baudry, A. Sené, I. A. Luk'yanchuk, L. Lahoche, J. F. Scott, Polarization vortex domains induced by switching electric field in ferroelectric films with circular electrodes. *Physical Review B* **90**, 024102 (2014).
14. W. J. Chen, Y. Zheng, B. Wang, Large and Tunable Polar-Toroidal Coupling in Ferroelectric Composite Nanowires toward Superior Electromechanical Responses. **5**, 11165 (2015).
15. Y. Tikhonov *et al.*, Controllable skyrmion chirality in ferroelectrics. *Sci Rep* **10**, 8657 (2020).

16. S. Yuan, Chen, W. J., Ma, L. L., Ye, J., Xiong, W. M., Liu, J. Y., Liu, Y. L., Wang, B., Zheng, Y., Defect-mediated vortex multiplication and annihilation in ferroelectrics and the feasibility of vortex switching by stress. *Acta Materialia* **148**, 330-343 (2018).
17. W. J. Chen, Yuan, S., Ma, L. L., Ye, J., Wang, B., Zheng, Y., Mechanical switching in ferroelectrics by shear stress and its implications on charged domain wall generation and vortex memory devices. *RSC Adv* **8**, 4434-4444 (2018).
18. P. Shafer *et al.*, Emergent chirality in the electric polarization texture of titanate superlattices. *Proceedings of the National Academy of Sciences*, (2018).
19. I. MacLaren *et al.*, On the origin of differential phase contrast at a locally charged and globally charge-compensated domain boundary in a polar-ordered material. *Ultramicroscopy* **154**, 57-63 (2015).
20. V. Grillo *et al.*, Highly efficient electron vortex beams generated by nanofabricated phase holograms. *Applied Physics Letters* **104**, 043109 (2014).
21. B. J. McMorran *et al.*, Electron Vortex Beams with High Quanta of Orbital Angular Momentum. *Science* **331**, 192 (2011).
22. J. Verbeeck, H. Tian, P. Schattschneider, Production and application of electron vortex beams. *Nature* **467**, 301-304 (2010).
23. M. Uchida, A. Tonomura, Generation of electron beams carrying orbital angular momentum. *Nature* **464**, 737-739 (2010).
24. M. W. Tate *et al.*, High Dynamic Range Pixel Array Detector for Scanning Transmission Electron Microscopy. *Microsc Microanal* **22**, 237-249 (2016).
25. A. K. Yadav *et al.*, Spatially resolved steady-state negative capacitance. *Nature* **565**, 468-471 (2019).
26. C. Kittel, *Introduction to Solid State Physics*. (Wiley, ed. 8, 2005).
27. R. D. King-Smith, D. Vanderbilt, Theory of polarization of crystalline solids. *Phys Rev B Condens Matter* **47**, 1651-1654 (1993).
28. A. Lubk, A. Beche, J. Verbeeck, Electron Microscopy of Probability Currents at Atomic Resolution. *Physical Review Letters* **115**, (2015).
29. K. X. Nguyen, Hovden, R., Tate, M. W., Purohit, P., Heron, J., Chang, C., Gruner, S. M., Muller, D. A. , in *Microsc Microanal*. (2015), vol. 21.
30. M. J. Humphry, B. Kraus, A. C. Hurst, A. M. Maiden, J. M. Rodenburg, Ptychographic electron microscopy using high-angle dark-field scattering for sub-nanometre resolution imaging. *Nat Commun* **3**, 730 (2012).
31. J. M. Rodenburg, R. H. T. Bates, The Theory of Super-Resolution Electron Microscopy Via Wigner-Distribution Deconvolution. *Philosophical Transactions of the Royal Society of London A: Mathematical, Physical and Engineering Sciences* **339**, 521-553 (1992).
32. Y. Jiang *et al.*, Electron ptychography of 2D materials to deep sub-angstrom resolution. *Nature* **559**, 343-349 (2018).
33. E. J. Kirkland, Advanced Computing in Electron Microscopy Second Edition Introduction. *Advanced Computing in Electron Microscopy, Second Ed*, 1-+ (2010).
34. I. Lazic, E. G. T. Bosch, S. Lazar, Phase contrast STEM for thin samples: Integrated differential phase contrast. *Ultramicroscopy* **160**, 265-280 (2016).
35. N. H. Dekkers, Lang, H. D. , Differential Phase-Contrast In A STEM. *Optik* **41**, 452-456 (1974).
36. A. Lubk, Zweck, J. , Differential phase contrast: An integral perspective. *Phys Rev A* **91**, (2015).
37. J. N. Chapman, Batson, P. E., Waddell, E. M., Ferrier, R. P. , The direct determination of magnetic domain wall profiles by differential phase contrast electron microscopy. *Ultramicroscopy* **3**, 203-214 (1978).
38. E. M. Waddell, Chapman, J. N. , Linear Imaging Of Strong Phase Objects Using Asymmetrical Detectors In STEM. *Optik* **54**, 84-96 (1979).
39. E. J. Kirkland, *Advanced Computing in Electron Microscopy*. (Plenum, NY, 1998).

40. A. R. Damodaran *et al.*, Phase coexistence and electric-field control of toroidal order in oxide superlattices. *Nat Mater* **16**, 1003-1009 (2017).
41. J. M. Zuo, J. C. H. Spence, *Electron Microdiffraction*. (Springer US, New York, ed. 1st, 1993).
42. E. M. Waddell, J. N. Chapman, Linear Imaging Of Strong Phase Objects Using Asymmetrical Detectors In STEM. *Optik* **54**, 83-96 (1979).
43. A. R. Damodaran *et al.*, Phase coexistence and electric-field control of toroidal order in oxide superlattices. *Nat Mater* **16**, 1003-1009 (2017).
44. J. C. Wojdeł, P. Hermet, M. P. Ljungberg, P. Ghosez, J. Íñiguez, First-principles model potentials for lattice-dynamical studies: general methodology and example of application to ferroic perovskite oxides. *J. Phys. Condens. Matter* **25**, 305401 (2013).

# Acknowledgments

**General**: The authors acknowledge discussions with Robert Hovden, Lena Fitting-Kourkoutis, and Megan E. Holtz. Microscopy facility support from John Grazul and Mariena Silvestry Ramos.

**Funding:** Y.J., D.A.M and ptychography supported by the U.S. Department of Energy, grant DE-SC0002334. Electron microscopy experiments by K.X.N. and equipment supported by the Cornell Center for Materials Research, through the National Science Foundation MRSEC program, award #DMR-1719875. Torque transfer theory (MC) supported by the Air Force Office of Scientific Research through the 2D Electronics MURI grant #FA9550-16-1-0031. Support for the MM-PAD development in SMG's lab was provided by the U.S. Department of Energy, grant DE-FG02-10ER46693. The adaptation to the STEM was supported by the Kavli Institute at Cornell for Nanoscale Science. J.I. acknowledges support from the Luxembourg National Research Fund under grant C15/MS/10458889 NEWALLS. P.G.F. and J.J. acknowledge financial support from the the Spanish Ministry of Science, Innovation and Universities through the grant No. PGC2018-096955-B. R.R. acknowledges support from the Quantum Materials program funded by the US Department of Energy, Office of Science.

**Author Contributions**: K.X.N. and D.A.M. designed the project. Electron microscopy and data analysis were carried out by K.X.N. and D.A.M.; PTO/STO superlattice samples grown by A.K.Y. under supervision of R.R. Electron microscope image simulation by Y.J., K.X.N. and M.C. Ptychography simulations by Y.J., supervised by D.A.M. Second-principles simulations were carried out by P.G.-F., P.A.-P., J.I. and J.J. Z. H. performed the phase-field simulations, supervised by L.Q.C. EMPAD electronics and design by M.W.T., P.P., supervised by S.M.G. All authors discussed the results and implications. K.X.N., J.J. R.R., L.W.M and D.A.M. wrote the paper.

**Competing interests:** Cornell has licensed the EMPAD to the FEI division of Thermo Fisher Scientific for commercial production. The authors declare that they have no competing financial interests. Correspondence and requests for materials should be addressed to D.A.M. (david.a.muller@cornell.edu) and K.X.N. (kn324@cornell.edu).

**Data and materials availability.** All relevant data are available from the corresponding author (kn324@cornell.edu, david.a.muller@cornell.edu) upon request. Code developed at Cornell, including visualization software for 4D data sets, is available from the corresponding author (kn324@cornell.edu, david.a.muller@cornell.edu) upon request.

# Figures

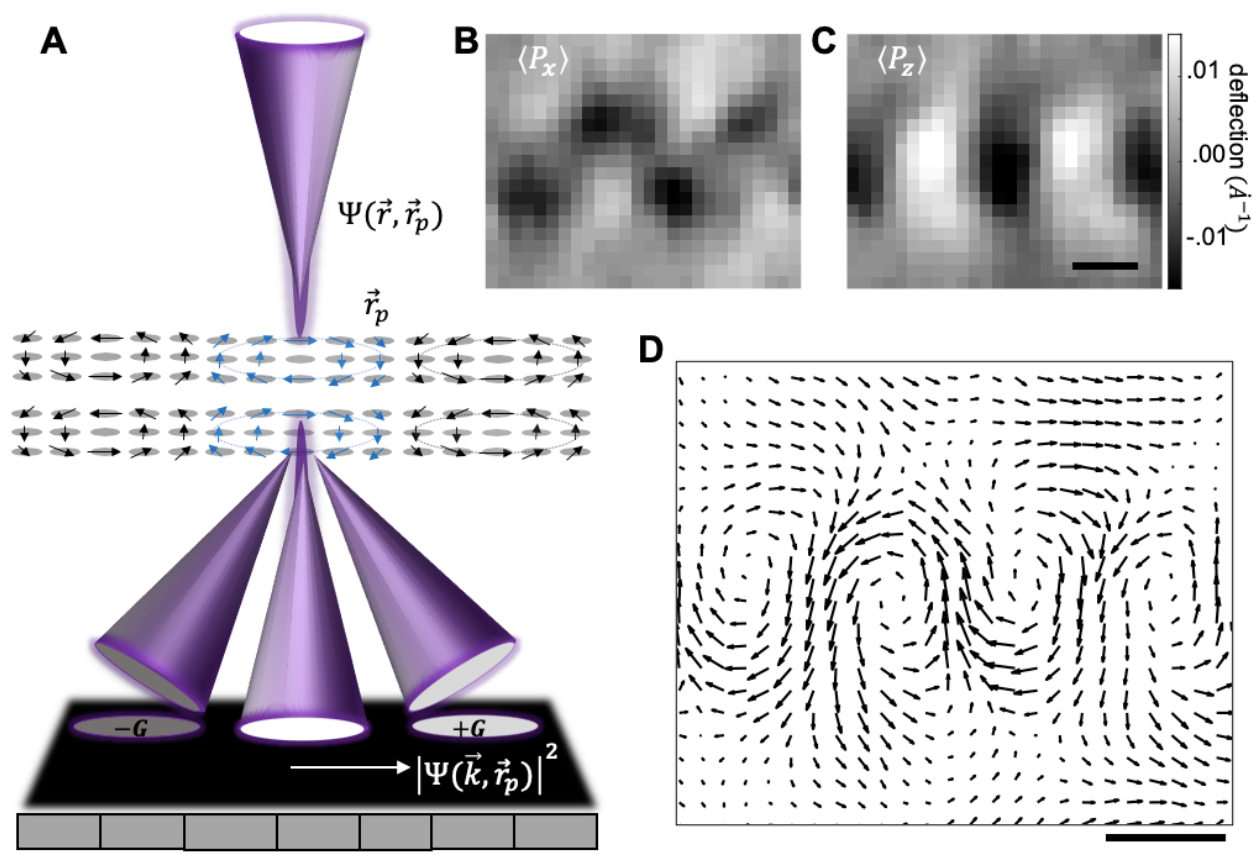

**Figure 1. Measuring complex polarization textures with EMPAD.** (**A**) Schematic of electron microscopy pixel array detector (EMPAD) placed in the diffraction plane, where a convergent beam electron diffraction (CBED) pattern is formed at the detector. Polarity causes an asymmetry in intensities of the conjugated pairs of diffracted disks at +**G** and -**G**, where **G** is the reciprocal lattice vector, indicated as light and dark gray disks. We utilize this aspect of the electron scattering distribution by taking the probability current flow, (**B**) $\langle P_x \rangle$ of [200] and (**C**) $\langle P_z \rangle$ of [020] diffracted disks, in units of inverse Angstrom, to reconstruct polarization vortices in (**D**). Scale bar in **B-D** represent a length of 2 nm.

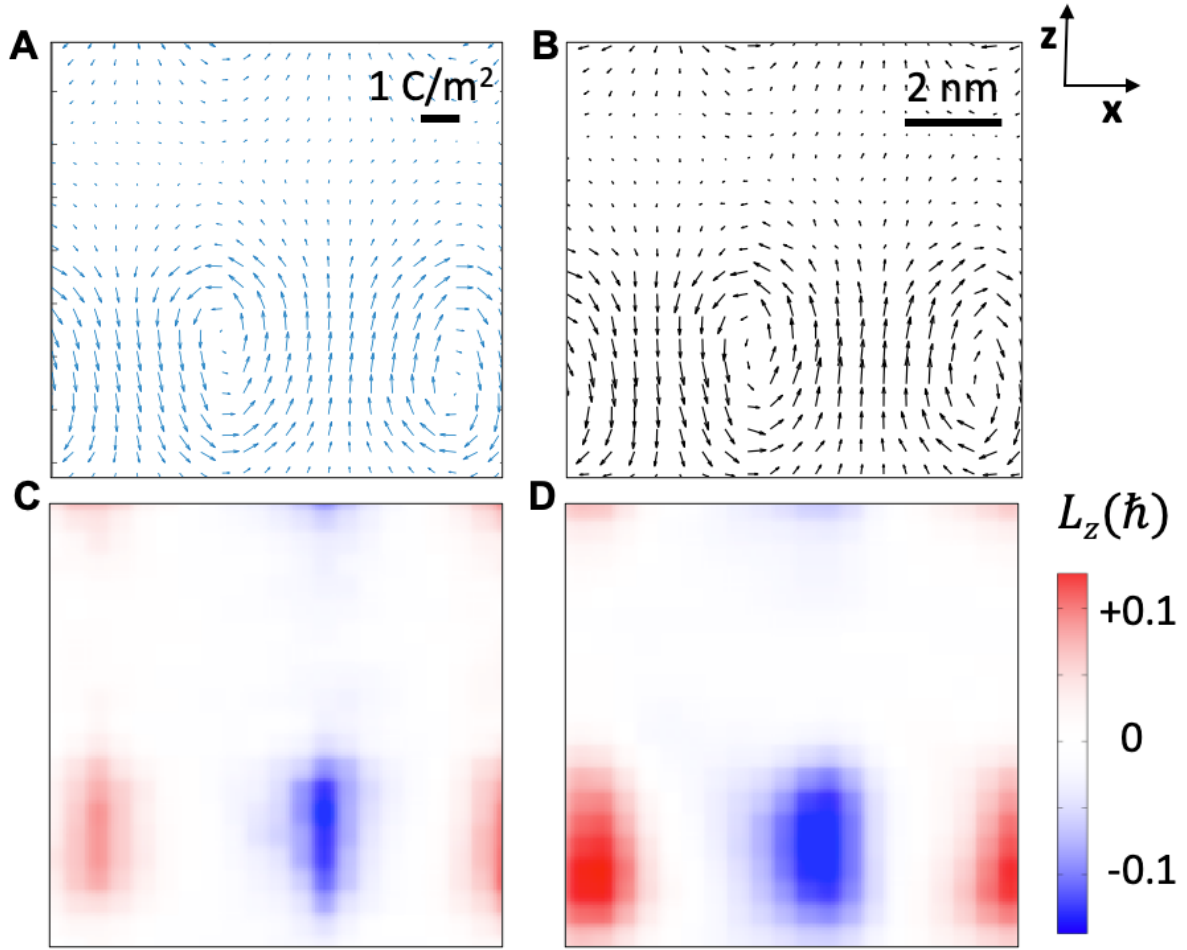

**Figure 2. Comparison of Orbital Angular Momentum and Torque Transfer Techniques.** Cross-sections along the vortex axis for a simulated 10×10 PbTiO$_3$/SrTiO$_3$ superlattice: (**A**) second-principles calculation of the polarization field. (**B**) Reconstructed vortices from $\langle p_x \rangle$ and $\langle p_z \rangle$ images of the (200) and $(\bar{2}00)$ diffracted disks calculated from the propagation of the electron beam through the simulated structure. (**C**) Change in orbital angular momentum reconstructed from the full wave function [Eq. (1)], and (**D**) from the integrated torque transfer from the electron to the sample calculated from the Fourier transform of Eq. (4) (see Method section). Both methods to estimate the transferred orbital angular momentum show very good agreement.

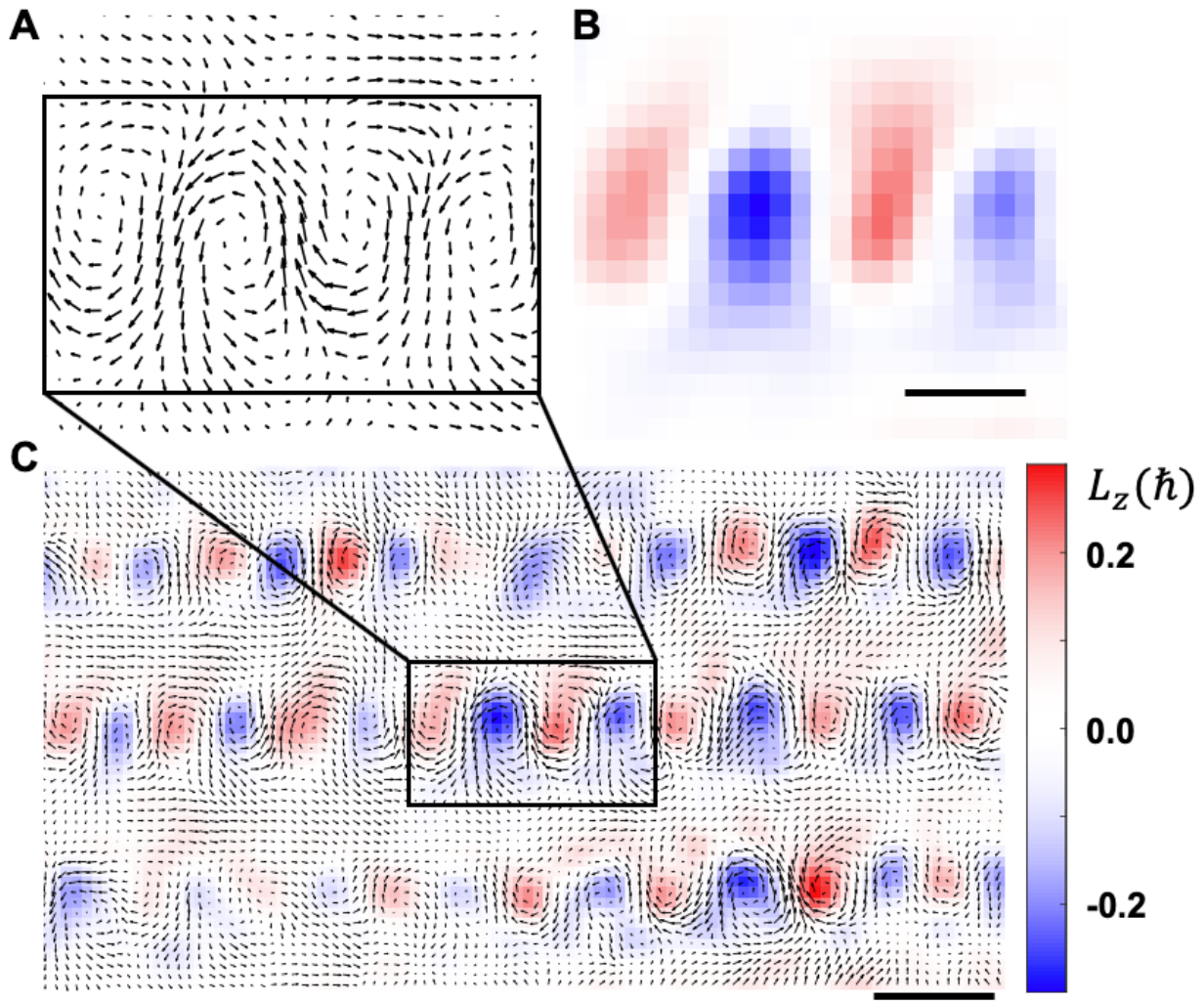

**Figure 3: Experimental Torque Transfer Results.** Using our analysis of polarity and analytical expression for orbital angular momentum (OAM), we can use our results to extract the OAM of the polarization vortices quantitatively. (**A**) Polarity vortices reconstructed from experimentally measured $\langle p_x \rangle$ and $\langle p_z \rangle$ along with (**B**) the measured torque transfer to electron beam for the same region. (**C**) Larger field of view of the sample showing torque transfer overlaid with the polarity map. Colorbar shows the change in angular momentum from the torque transfer in units of ℏ. Black scale bar in (**B**) is 2 nanometers and (**C**) is 5 nanometers.

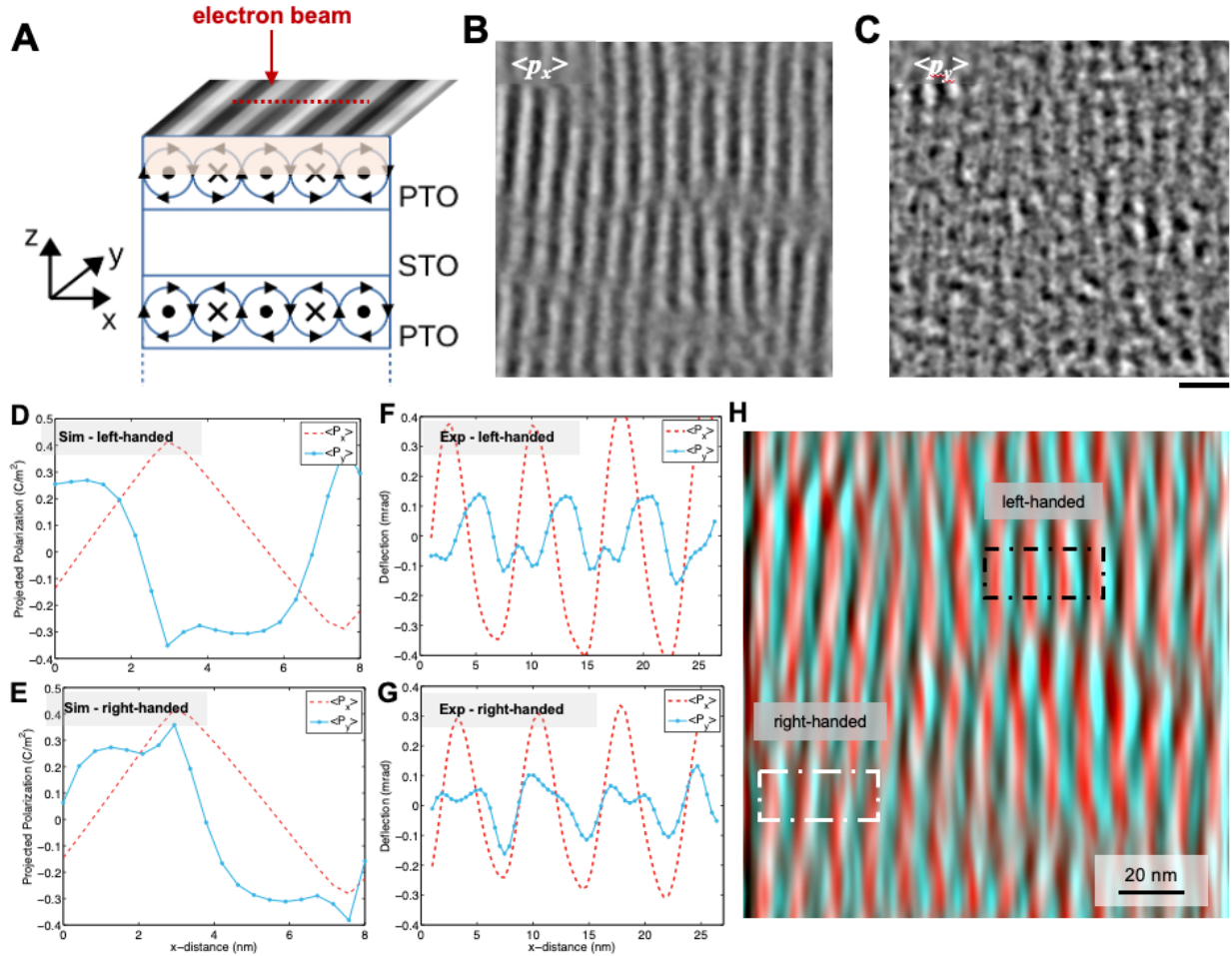

**Figure 4: Solving 3-dimensional chirality.** Plane view imaging of the polarization textures. The underlying polarization texture relative to the electron microscope image is sketched in (**A**), where vortices in the superlattice are represented as arrows describing circles in the (*x*,*z*) plane plus dots and crosses indicating polarization pointing along the positive or negative *y* axis. Here, the top half of the vortices is highlighted in red representing the fact that the probability current signal mostly comes from the top half of the vortices (see text). Experimental images of (**B**) $<p_x>$ and (**C**) $<p_y>$ images from a 12×12 superlattice using a 1.4 mrad semi-converged angle probe at 300 keV and low-pass filtered to the probe's Nyquist limit. We observed that the stripes have higher contrast in $<p_x>$ than in $<p_y>$. Black scale bar under (**C**) represents 20 nm. Second-principles calculation of the polarization field show projected polarization of $<p_x>$ and $<p_y>$ as line profiles for a (**D**) left-handed chiral structure and the (**E**) right-handed chiral structure. To visualize the registration between components, we superimpose and bandpass filter our results from (**B**) $<p_x>$ and (**C**) $<p_y>$ as a false–color image with $<p_x>$ and $<p_y>$ in the red and blue/green channels respectively (**H**) so that pale or white colors indicate a superposition of peaks in *x* and *y*. Line profiles taken from two different regions (within the dot-dashed rectangles) in (**H**) show both (**F**) left-handed and (**G**) right-handed chirality.

# Supplementary Materials

**Supplementary Figures**

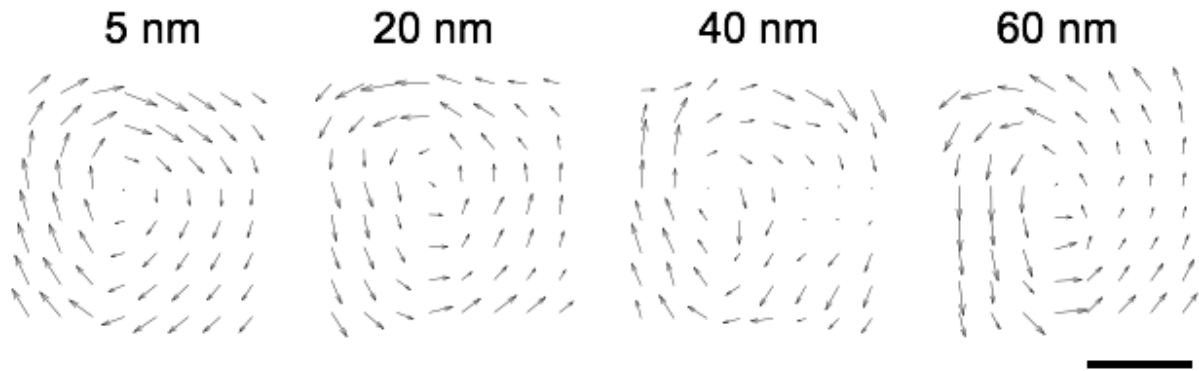

**Fig. S1. Polarization vortices orientation at different sample thickness.** Reconstructed vortices from $\langle p_x \rangle$ and $\langle p_y \rangle$ images of the $(200)$ and $(\bar{2}00)$ diffracted disks calculated from a simulated structure at thicknesses of 5, 20, 40 and 50 nanometers. Here, we observed that channeling of the electron beam causes the polarization direction to switch depending on the thickness with a period of 20 nanometers, reflecting contrast reversals in the underlying {200} diffraction peaks from dynamical scattering in the sample. Black scale bar represents 3 nanometers.

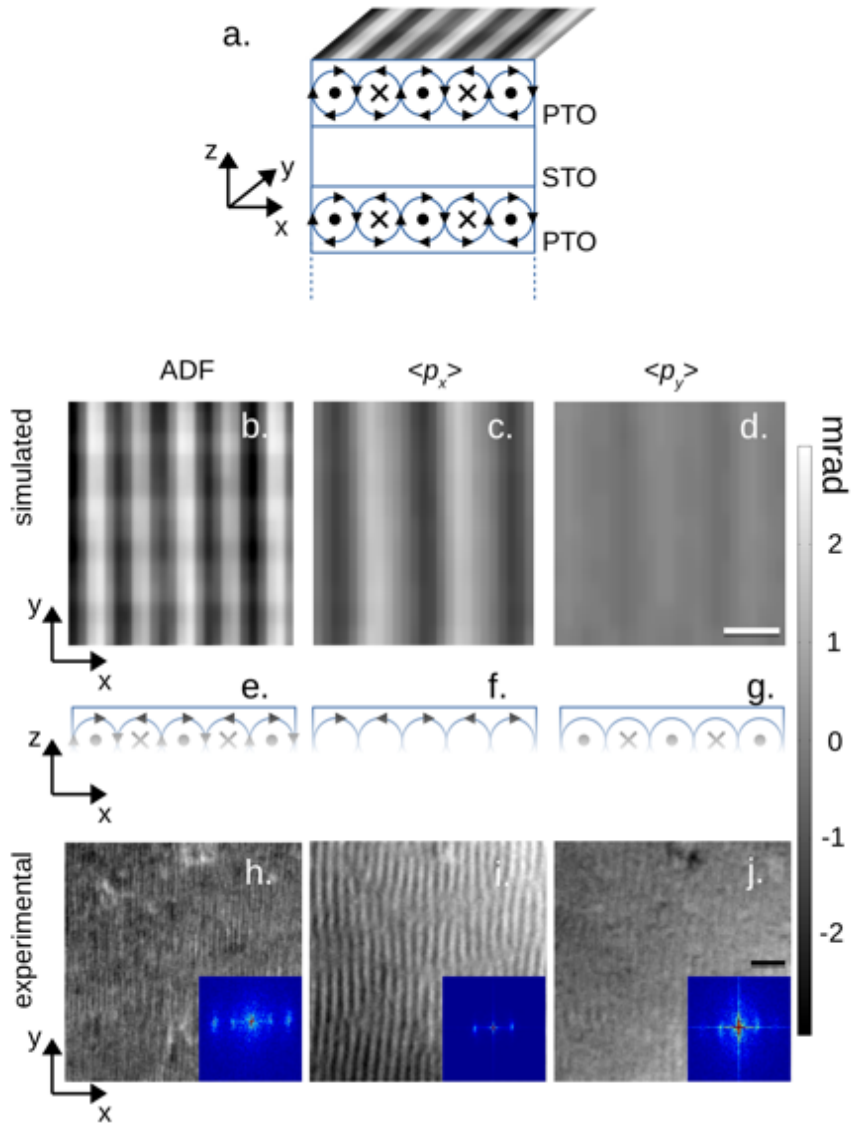

**Fig. S2. Plane view imaging of the polarization textures.** The underlying polarization texture relative to the electron microscope image is sketched (a), where vortices in the superlattice are represented as arrows describing circles in the (*x,z*) plane plus dots and crosses indicating polarization pointing along the positive or negative *y* axis. (b) Annular dark field (ADF), (c) $<p_x>$ and (d) $<p_y>$ images reconstructed from the coordinates of a 10×10 superlattice using 1.76 mrad semi-converged angle probe at 300 keV. (e-g) Show the features of the polarization texture for which the measurements in the panels above and below are sensitive. The fading of the sketches in (e-g) represent the fact that the probability current signal comes mostly from the top half of the vortices (see text). Experimental results from 12×12 superlattice using the same imaging parameters as simulation for (h) ADF, (i) $<p_x>$ and (j) $<p_y>$. By looking at a larger field of view in (h) ADF, and probability current flow (i) $<p_x>$ and (j) $<p_y>$ we observed that the stripes have higher contrast in $<p_x>$ than in $<p_y>$, although faint contrast is seen in $<p_y>$. Fourier transforms (FT) of (h), (i), and (j) are represented as insets to each figure respectively. Here, we observed double periodicity in the FT (g), while the FT (i) and (j) show single periodicity with the (i) FT of $<p_x>$ having 6 times more intensity than the (j) FT of $<p_y>$. White scale bar in (d) represents 5 nm. Black scale bar under (j) represents 30 nm.

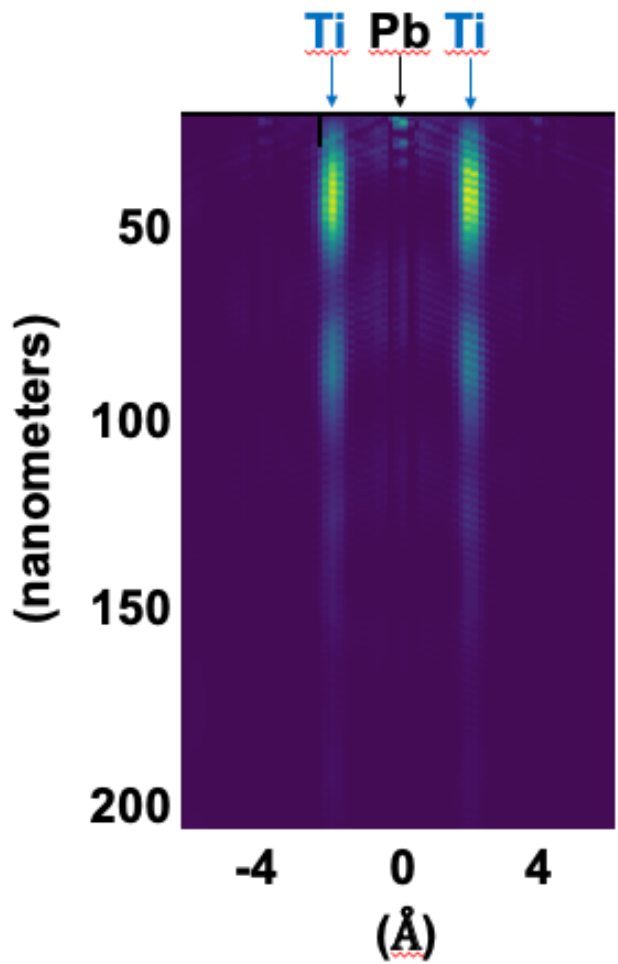

**Fig. S3. Map of probe intensity and its dependence on thickness.** Using an electron beam with a 1.76 mrad convergence angle at 300 keV Multislice simulation of probe intensity from a 6×6 PbTiO$_3$/SrTiO$_3$ structure, where the electron probe is shown channeling down the [001] zone axis. Multislice conditions are the same as experimental conditions used to those recorded with the EMPAD on a FEI Titan Themis.

# Supplementary Text

## Measuring Polarity and Torque Transfer from Probability Current Images

Here we derive the central relationship that connects the measured probability current flow of an electron beam with wave function $\Psi(\vec{r}, \vec{r}_p)$ centered about probe position $\vec{r}_p$ to the torque transfer using the strong phase approximation. The starting point is the measurement of the center of mass image (CoM), $\langle \vec{p} \rangle$ formed by scanning the probe position $\vec{r}_p$ and measuring the angular distribution of scattered electron beam at each probe position, $|\Psi(\vec{k}, \vec{r}_p)|^2$, where $\vec{k}$ and $\vec{r}$ are the conjugate variables in the back focal plane and image plane respectively [1,2]. A center of mass (COM) image has each pixel value equal to the centroid of the associated diffraction pattern where

$$\langle \vec{p}(\vec{r}_p) \rangle = \int \hbar \vec{k} |\Psi(\vec{k}, \vec{r}_p)|^2 d\vec{k} \tag{S1}$$

which follows from the definition of $\langle \vec{p} \rangle$ written out in a momentum basis [1,2].

Expanding $\langle \vec{p} \rangle = \langle \Psi | \hat{p} | \Psi \rangle$ in a position basis and using $\hat{p} = -i\hbar \vec{\nabla}$ we obtain

$$\langle \vec{p}(\vec{r}_p) \rangle = -\hbar i \int \Psi^*(\vec{r}, \vec{r}_p) \vec{\nabla} \Psi(\vec{r}, \vec{r}_p) d\vec{r} \tag{S2}$$

Taking into account that the momentum operator is an Hermitian (i. e. self-adjoint operator), then

$$\langle \vec{p}(\vec{r}_p) \rangle = \hbar i \int \Psi(\vec{r}, \vec{r}_p) \vec{\nabla} \Psi^*(\vec{r}, \vec{r}_p) d\vec{r}. \tag{S3}$$

Adding together the two last Equations, we arrive to the conclusion that $\langle \vec{p} \rangle$ differs from the expectation value of probability current flow $\langle \vec{j} \rangle$ by a factor of twice the electron's mass, $m$ [1,2]:

$$\langle \vec{p}(\vec{r}_p) \rangle = \frac{\hbar}{2i} \int \Psi^*(\vec{r}, \vec{r}_p) \vec{\nabla} \Psi(\vec{r}, \vec{r}_p) - \Psi(\vec{r}, \vec{r}_p) \vec{\nabla} \Psi^*(\vec{r}, \vec{r}_p) \, d\vec{r}$$
$$= 2m \langle \vec{j} \rangle \tag{S4}.$$

From Equation S4, we find that the expectation value of momentum or center of mass (CoM), $\langle \vec{p} \rangle$, can be used interchangeably with the probability current flow, $\langle \vec{j} \rangle$. Furthermore, this is the classical definition of a current, where the rate of flow from an electric charge is related to the net flow of electron beam in the sample measured as $\langle \vec{p} \rangle$.

In **Section 1**, we review the already-derived connection between the center of mass images and the gradient of the potential to establish a consistent notation. In **Section 2**, we derive the new relationship between torque transfer and the center of mass images. In **Section 3 and Section 4**, we show how polarity is encoded in the probability current flow to a pair of conjugate diffracted beams.

### Section 1: Relation between Center of Mass and Potential Gradient Images

To connect $\langle \vec{p} \rangle$ of the exit wave to the scattering potential, we next make the strong phase approximation which should hold so long as the probe amplitude does not change dramatically in sample. In the strong phase approximation, the exit wavefunction is just a product of the initial wavefunction, $\Psi_0$, and a phase term from the sample potential, $V(\vec{r})$.

$$\Psi(\vec{r}, \vec{r}_p) = exp(i\sigma V(\vec{r})) \Psi_0(\vec{r} - \vec{r}_p), \tag{S5}$$

where $\sigma$ is the usual interaction parameter [3]. Substituting eqn. (S5) into eqn. (S2) gives:

$$\langle \vec{p}(\vec{r}_p) \rangle = -i\hbar \int \left[ i\sigma \vec{\nabla} V(\vec{r}) \Psi_0(\vec{r} - \vec{r}_p) + \vec{\nabla} \Psi_0(\vec{r} - \vec{r}_p) \right] \exp(i\sigma V(\vec{r})) \exp(-i\sigma V(\vec{r})) \Psi_0^*(\vec{r} - \vec{r}_p) \, d\vec{r} \tag{S6}$$

$$\langle \vec{p}(\vec{r}_p) \rangle = -i\hbar \int \Psi_0^*(\vec{r} - \vec{r}_p) \vec{\nabla} \Psi_0(\vec{r} - \vec{r}_p) d\vec{r} + \hbar \sigma \int \vec{\nabla} V(\vec{r}) |\Psi_0(\vec{r} - \vec{r}_p)|^2 d\vec{r}$$

If the beam is symmetric [4], then the first term goes to zero and the second term can be written as a convolution:

$$\langle \vec{p}(\vec{r}_p) \rangle = \hbar \sigma |\Psi_0(\vec{r}_p)|^2 \otimes \vec{\nabla} V(\vec{r}_p) \tag{S7}$$

which establishes the conditions under which the center of mass image is a convolution of the potential gradient and the probe. If the beam is not symmetric, then the first term, which depends only on the incident beam, provides a constant offset uniform background which can be subtracted off, and the second term changes equation S7 to a cross correlation [5].

**Section 2: Calculating Torque from COM Images**

To calculate torque, we start with a relation in Ehrenfest's theorem that connects $\langle \vec{p} \rangle$ to the gradient of the projected potential [1, 2]. Second, we find that the torque operator for an electron in the beam can be defined as

$$\hat{\Gamma} = \hat{\vec{r}} \times -\vec{\nabla} \hat{V} \tag{S8}$$

We can calculate its expectation value by expanding in the position basis.

$$\langle \hat{\Gamma} \rangle = \langle \Psi | \hat{\Gamma} | \Psi \rangle \tag{S9}$$

and the $z$-component is

$$\langle \hat{\Gamma}_z \rangle = \begin{aligned} &\int \langle \Psi | \vec{r}'' \rangle \langle \vec{r}'' | \hat{x} | \vec{r}' \rangle \left\langle \vec{r}' \left| -\frac{\partial \hat{V}}{\partial y} \right| \vec{r} \right\rangle \langle \vec{r} | \Psi \rangle d\vec{r} d\vec{r}' d\vec{r}'' \\ &+ \int \langle \Psi | \vec{r}'' \rangle \langle \vec{r}'' | \hat{y} | \vec{r}' \rangle \left\langle \vec{r}' \left| \frac{\partial \hat{V}}{\partial x} \right| \vec{r} \right\rangle \langle \vec{r} | \Psi \rangle d\vec{r} d\vec{r}' d\vec{r}'' \end{aligned} \tag{S10}$$

Since the position basis are eigenvectors of the position and potential operators, we can expand the integral

$$\langle \hat{\Gamma}_z \rangle = \int x \frac{-\partial V}{\partial y}(\vec{r}') \Psi^*(\vec{r}'') \Psi(\vec{r}) \delta(\vec{r}'' - \vec{r}') \delta(\vec{r}' - \vec{r}) d\vec{r} d\vec{r}' d\vec{r}'' + \int y \frac{\partial V}{\partial x}(\vec{r}') \Psi^*(\vec{r}'') \Psi(\vec{r}) \delta(\vec{r}'' - \vec{r}') \delta(\vec{r}' - \vec{r}) d\vec{r} d\vec{r}' d\vec{r}'' \tag{S11}$$

We only show the calculation for one of the terms in the sum. Both are similar, and combining gives

$$\langle \hat{\Gamma}_z \rangle = \int x \frac{-\partial V}{\partial y} |\Psi(\vec{r})|^2 d\vec{r} + \int y \frac{\partial V}{\partial x} |\Psi(\vec{r})|^2 d\vec{r} \tag{S12}$$

This is the general formula where $\Psi(\vec{r})$ is the exit wavefunction. In the context of a scanning beam, we can write our exit wavefunction as the probe shifted to the relevant scan point $\Psi(\vec{r} - \vec{r}_p)$. We also want to measure with respect to $\vec{r}_p$ as the origin. Therefore, we re-write for the torque in the $z$ direction as

$$\Gamma_z(\vec{r}_p) = \int [(\vec{r} - \vec{r}_p) \times (-\vec{\nabla} V(\vec{r}))]_z |\Psi(\vec{r} - \vec{r}_p)|^2 d\vec{r} \tag{S13}$$

where $V(\vec{r})$ is the potential of the specimen.

We take advantage of the symmetry of the probe to re-write this equation to look like a convolution (for asymmetric probes this will remain a cross-correlation, a result which will carry through without loss of generality):

$$\Gamma_z(\vec{r}_p) = \int [(\vec{r}_p - \vec{r}) \times -\vec{\nabla} V(\vec{r})]_z |\Psi(\vec{r}_p - \vec{r})|^2 d\vec{r} \tag{S14}$$

Here, we have introduced Eq. (S13) and (S14) in vector form, from which we could relate this back to Eq. (S12). Furthermore, convolutions in real space are multiplications in Fourier space (equation A3 in Appendix) which by taking a Fourier transform gives

$$\mathcal{F}[\Gamma_z(\vec{r})] = -\mathcal{F}\left[\frac{-\partial V}{\partial y}(\vec{r})\right] \mathcal{F}[x|\Psi(\vec{r})|^2] + \mathcal{F}\left[\frac{\partial V}{\partial x}(\vec{r})\right] \mathcal{F}[y|\Psi(\vec{r})|^2] \tag{S15}$$

At this point, we need to find the gradient of the potential. As shown in equation S7, that by using the strong phase approximation, our COM images are related to the gradient of the potential by a convolution. For example, the COM in the $x$ direction is with $\Psi_0$ being the incident wavefunction given by

$$\langle p_x(\vec{r}_p)\rangle = \hbar\sigma |\Psi_0(\vec{r}_p)|^2 \otimes \frac{\partial V(\vec{r}_p)}{\partial x} \tag{S16}$$

Taking a Fourier transform gives

$$\mathcal{F}[\langle p_x(\vec{r}_p)\rangle] = \hbar\sigma \, \mathcal{F}[|\Psi(\vec{r})|^2] \, \mathcal{F}\left[\frac{\partial V}{\partial x}(\vec{r})\right] \tag{S17}$$

We can now write Eq. (S15) as

$$\Gamma_z(\vec{r}) = \mathcal{F}^{-1}\left\{\frac{-\mathcal{F}[\langle p_y(\vec{r})\rangle]\mathcal{F}[x|\Psi(\vec{r})|^2] + \mathcal{F}[\langle p_x(\vec{r})\rangle]\mathcal{F}[y|\Psi(\vec{r})|^2]}{\hbar\sigma\mathcal{F}[|\Psi(\vec{r})|^2]}\right\} \tag{S18}$$

This is the desired result, describing the torque in terms of only the experimentally-measured quantities $\langle p_x(\vec{r}_p)\rangle$, $\langle p_y(\vec{r}_p)\rangle$, $|\Psi_0(\vec{r})|^2$. A key observation here is that in the strong phase approximation (equation S4), $|\Psi(\vec{r})|^2 = |\Psi_0(\vec{r})|^2$ and the incident beam shape can be measured directly at medium resolution, or with the aberration-correction software at high-resolution. At medium spatial resolution (non-overlapping disks), multislice simulations indicate this approximation is robust for sample thicknesses up to 20 nm at 300 keV.

Now $\mathcal{F}[|\Psi(\vec{r})|^2]$ is peaked at zero frequency and is zero at frequencies with k-vectors magnitudes larger than the diameter of the probe-forming aperture. To avoid dividing by zero or by values arbitrarily close to zero in $\mathcal{F}[|\Psi(\vec{r})|^2]$, we pass this through a low-pass filter to suppress the high frequency noise beyond the aperture cutoff. Equation S18 can be trivially modified to incorporate an optimal Wiener filter, recognizing that eqn. (S18) is essentially deconvolving the effect of the probe contrast transfer function (CTF) from the torque measurement. Omitting the division by the probe CTF leaves us with the torque measurement blurred out to probe resolution. This may be preferable for noisy data or thick samples where the strong phase approximation may no longer hold.

## Section 3: Measuring Polarity from the CoM image in the Strong Phase Approximation

Polarity is usually calculated within a Bloch-wave formalism [6] to account for the multiple scattering of the electron beam through the sample. While there are analytic results for special cases, that approach is more useful computationally than for insight. Here we consider a thin sample described by the strong phase approximation that provides a simpler result, useful for understanding the contributions to the polarity measurement, while still retaining the key symmetries of the more complicated theory.

Starting with equation (S4), and as shown by Deb. and coworkers in Ref. [7], the diffraction pattern expanded out to third order is:

$$|\Psi(\vec{k},\vec{r}_p)|^2 = \begin{aligned} &|\Psi_0(\vec{k})|^2 + \frac{\sigma}{\pi}\text{Im}\{\Psi_0(\vec{k},\vec{r}_p)[\Psi_0^*(\vec{k},\vec{r}_p)\otimes V^*(\vec{k})]\} \\ &+ \frac{\sigma^2}{4\pi^2}\{|\Psi_0(\vec{k},\vec{r}_p)\otimes V(\vec{k})|^2 - \text{Re}\left[\Psi_0(\vec{k},\vec{r}_p)\left(\Psi_0^*(\vec{k},\vec{r}_p)\otimes V_2^*(\vec{k})\right)\right]\} \\ &+ \frac{\sigma^3}{8\pi^3}\left\{\begin{array}{l}\text{Im}\left[\left(\Psi_0(\vec{k},\vec{r}_p)\otimes V(\vec{k})\right)\left(\Psi_0^*(\vec{k},\vec{r}_p)\otimes V_2^*(\vec{k})\right)\right] \\ -\frac{1}{3}\text{Im}\left[\Psi_0(\vec{k},\vec{r}_p)\left(\Psi_0^*(\vec{k},\vec{r}_p)\otimes V_3^*(\vec{k})\right)\right]\end{array}\right\} \\ &+ \cdots \end{aligned} \qquad (S19)$$

where

$$\Psi_0(\vec{k},\vec{r}_p) = \Psi_0(\vec{k})\exp(-i\vec{k}\cdot\vec{r}_p), \qquad (S20)$$

$$V_2(\vec{k}) = V(\vec{k})\otimes V(\vec{k}), \qquad (S21)$$

$$V_3(\vec{k}) = V(\vec{k})\otimes V(\vec{k})\otimes V(\vec{k}). \qquad (S22)$$

The only term outside the bright disk field that is asymmetric and therefore contributes to the Center-of-Mass signal from the diffracted beams is the third order term

$$\text{Im}\left[\left(\Psi_0(\vec{k},\vec{r}_p)\otimes V(\vec{k})\right)\left(\Psi_0^*(\vec{k},\vec{r}_p)\otimes V_2^*(\vec{k})\right)\right]. \qquad (S23)$$

We assume our sample is a crystal and can therefore write our crystal potential in reciprocal space as

$$V(\vec{k}) = \mathcal{F}[V(\vec{r})] = \sum_{\vec{G}} U_{\vec{G}}\exp(i\phi_{\vec{G}})\delta(\vec{k}-\vec{G}), \qquad (S24)$$

where $\vec{G} = h\vec{g}_1 + k\vec{g}_2 + l\vec{g}_3$ is a reciprocal lattice vector and $U_{\vec{G}}$ and $\phi_{\vec{G}}$ are real. Substituting $V(\vec{k})$ from equation (S24) into equation (S23) introduces 3 sums over reciprocal lattice vectors labelled $\vec{G}_1, \vec{G}_2, \vec{G}_3$, and after some algebra, equation (S23) becomes

$$\begin{aligned}&\text{Im}\left[\left(\Psi_0(\vec{k},\vec{r}_p)\otimes V(\vec{k})\right)\left(\Psi_0^*(\vec{k},\vec{r}_p)\otimes V_2^*(\vec{k})\right)\right] \\ &= \sum_{\vec{G}_1,\vec{G}_2,\vec{G}_3}\Psi_0(\vec{k}-\vec{G}_1)\Psi_0^*(\vec{k}-\vec{G}_2-\vec{G}_3)U_{\vec{G}_1}U_{\vec{G}_2}U_{\vec{G}_3}\sin\left(\phi_{\vec{G}_1}-\phi_{\vec{G}_2}-\phi_{\vec{G}_3}\right)\end{aligned} \qquad (S25).$$

Equation (S25) gives rise to the polarity term and the phase portion in equation (S25) is recognized as the three-phase invariant of crystallography:

$$\phi = \phi_{\vec{G}_1} - \phi_{\vec{G}_2} - \phi_{\vec{G}_1-\vec{G}_2}. \qquad (S26)$$

which is invariant under a change of origin in real space: A shift in the origin by $\vec{r}_0$ leads to a phase shift $\phi'_{\vec{G}} = \phi_{\vec{G}} + \vec{G} \cdot \vec{r}_0$, but as the vectors $\vec{G}_1 - \vec{G}_2 - \vec{G}_1 + \vec{G}_2 = 0$, the offset $(\vec{G}_1 - \vec{G}_2 - \vec{G}_1 + \vec{G}_2) \cdot \vec{r}_0 = 0$ as well, thus there is no phase shift when choosing a new origin, making $\sin(\phi)$ a good metric for tracking components of the polar order parameter.

**Section 4: Polarity from the Center of Mass of Conjugate Diffraction Spots**

Polarity can be sensed most simply through the use of Friedel pairs at $\vec{G}_1$ and $-\vec{G}_1$. Focusing on the diffraction spot centered around $\vec{G}_1$, we remove the $\vec{G}_1$ summation from equation (S25), and similarly for $-\vec{G}_1$. The probability current flow given by the CoM measurement will then become the sum of the first moments of the $\vec{G}_1$ and $-\vec{G}_1$ spots:

$$\int \vec{k} \sum_{\vec{G}_2} |\Psi_0(\vec{k} - \vec{G}_1)|^2 U_{\vec{G}_1} U_{\vec{G}_2} U_{\vec{G}_1 - \vec{G}_2} \sin\left(\phi_{\vec{G}_1} - \phi_{\vec{G}_2} - \phi_{\vec{G}_1 - \vec{G}_2}\right) d\vec{k}$$
$$+ \int \vec{k} \sum_{\vec{G}_3} |\Psi_0(\vec{k} + \vec{G}_1)|^2 U_{-\vec{G}_1} U_{\vec{G}_3} U_{\vec{G}_1 - \vec{G}_3} \sin\left(\phi_{-\vec{G}_1} - \phi_{\vec{G}_3} - \phi_{\vec{G}_1 - \vec{G}_3}\right) d\vec{k} \tag{S26}$$

where we have separated the integral for later convenience. We can make a variable substitution for the integration variables $\vec{k}$ so that they are centered around the diffraction spot:

$$\sum_{\vec{G}_2} U_{\vec{G}_1} U_{\vec{G}_2} U_{\vec{G}_1 - \vec{G}_2} \sin\left(\phi_{\vec{G}_1} - \phi_{\vec{G}_2} - \phi_{\vec{G}_1 - \vec{G}_2}\right) \int (\vec{k} + \vec{G}_1) |\Psi_0(\vec{k})|^2 d\vec{k}$$
$$+ \sum_{\vec{G}_3} U_{-\vec{G}_1} U_{\vec{G}_3} U_{\vec{G}_1 - \vec{G}_3} \sin\left(\phi_{-\vec{G}_1} - \phi_{\vec{G}_3} - \phi_{\vec{G}_1 - \vec{G}_3}\right) \int (\vec{k} - \vec{G}_1) |\Psi_0(\vec{k})|^2 d\vec{k} \tag{S27}$$

The first part of each integral $\int \vec{k} |\Psi_0(\vec{k})|^2 d\vec{k}$ is just the first moment of the un-scattered bright-field disk, which is zero for a non-aberrated or symmetric incident beam. The second part is just the total intensity of the beam $I_0$ times the reciprocal lattice vector. This simplifies the expression to

$$\vec{G}_1 I_0 \sum_{\vec{G}_2} U_{\vec{G}_1} U_{\vec{G}_2} U_{\vec{G}_1 - \vec{G}_2} \sin\left(\phi_{\vec{G}_1} - \phi_{\vec{G}_2} - \phi_{\vec{G}_1 - \vec{G}_2}\right)$$
$$- \vec{G}_1 I_0 \sum_{\vec{G}_3} U_{-\vec{G}_1} U_{\vec{G}_3} U_{\vec{G}_1 - \vec{G}_3} \sin\left(\phi_{-\vec{G}_1} - \phi_{\vec{G}_3} - \phi_{\vec{G}_1 - \vec{G}_3}\right) \tag{S28}$$

We can re-index $\vec{G}_3 = -\vec{G}_2$ without changing the second summation:

$$\vec{G}_1 I_0 \sum_{\vec{G}_2} \begin{aligned} & U_{\vec{G}_1} U_{\vec{G}_2} U_{\vec{G}_1 - \vec{G}_2} \sin\left(\phi_{\vec{G}_1} - \phi_{\vec{G}_2} - \phi_{\vec{G}_1 - \vec{G}_2}\right) \\ & - U_{-\vec{G}_1} U_{-\vec{G}_2} U_{-\vec{G}_1 + \vec{G}_2} \sin\left(\phi_{-\vec{G}_1} - \phi_{-\vec{G}_2} - \phi_{-\vec{G}_1 + \vec{G}_2}\right) \end{aligned} \tag{S29}$$

For a real potential $V(\vec{r})$, $V^*(\vec{k}) = V(-\vec{k})$. This means $U_{\vec{G}} = U_{-\vec{G}}$ and $\phi_{\vec{G}} = -\phi_{-\vec{G}}$ which gives our final expression for the Center of mass of the Friedel pair:

$$\langle \vec{p}_{\{\vec{G}_1\}} \rangle = \left(\frac{\hbar \sigma^3}{8\pi^3}\right) 2\vec{G}_1 I_0 \sum_{\vec{G}_2} U_{\vec{G}_1} U_{\vec{G}_2} U_{\vec{G}_1 - \vec{G}_2} \sin(\phi) \tag{S30}$$

where three-phase invariant $\phi = \phi_{\vec{G}_1} - \phi_{\vec{G}_2} - \phi_{\vec{G}_1 - \vec{G}_2}$ as before. However, if the crystal is also centrosymmetric, i.e. $V(\vec{r}) = V(-\vec{r})$, then the Fourier transform is also pure real, so all $\phi_{\vec{G}} = 0$ which means equation (S3.13) simplifies to 0 for non-polar materials. This result remains 0 even under a shift in origin that breaks the even symmetry of the crystal potential $V(\vec{r})$ thanks to the three-phase invariant. More generally, as $\sin(\phi)$ is independent of the choice of origin in real space, it can be used as a good order parameter for describing the component of the polarity along $\vec{G}_1$.

Finally, the Fourier coefficients of the potential, $U_{\vec{G}}$, are sensitive to both the nuclear and electronic contributions of the total potential for electron scattering. In other words the measured dipole density probed is net/total dipole density.

**Appendix: Fourier Convention**

Fourier Transform convention:

$$F(\vec{k}) = \frac{1}{2\pi} \int f(\vec{r}) \exp(-i\vec{k} \cdot \vec{r}) d\vec{r} \tag{A1}$$

$$f(\vec{r}) = \frac{1}{2\pi} \int F(\vec{k}) \exp(i\vec{k} \cdot \vec{r}) d\vec{k} \tag{A2}$$

When using this convention, we get these properties:

$$\mathcal{F}[A(\vec{r})B(\vec{r})] = \frac{1}{2\pi} A(\vec{k}) \otimes B(\vec{k}) \tag{A3}$$

$$\mathcal{F}[A(\vec{r}) \otimes B(\vec{r})] = 2\pi A(\vec{k}) B(\vec{k}) \tag{A4}$$

$$\mathcal{F}\left[\frac{\partial^n}{\partial x_i^n} A(\vec{r})\right] = (ik_i)^n A(\vec{k}) \tag{A5}$$

where $\otimes$ represents the convolution operator.

**References**

[1] A Lubk and J Zweck. Differential phase contrast: An integral perspective. *Physical Review A*, 91(2):023805, 2015.

[2] K Müller, Florian F Krause, Armand Beche, Marco Schowalter, Vincent Galioit, Stefan Löffler, Johan Verbeeck, Josef Zweck, Peter Schattschneider, and Andreas Rosenauer. Atomic electric fields revealed by a quantum mechanical approach to electron picodiffraction. *Nature communications*, 5, 2014.

[3] Earl J. Kirkland, Advanced Computing in Electron Microscopy. 2nd edn, (Springer US, 2010).

[4] Michael C. Cao, Yimo Han, Zhen Chen, Yi Jiang, Kayla X. Nguyen, Emrah Turgut, Gregory D. Fuchs, David A. Muller, Theory and Practice of Electron Diffraction from Single Atoms and Extended Objects Using an EMPAD. Microscopy, 67: i150 – i161, 2018.

[5] Ivan Lazic, Eric GT Bosch, and Sorin Lazar. Phase contrast stem for thin samples: Integrated differential phase contrast. *Ultramicroscopy*, 160: 265– 280, 2016.

[6] Zuo, J. M. & Spence, J. C. H. *Electron Microdiffraction*. 1st edn, (Springer US, 1993).

[7] P Deb, Y Han, ME Holtz, J Park, DA Muller. Imaging Polarity in Two Dimensional Materials by Breaking Friedel's Law. *Ultramicroscopy*, 215:1-9, 2020.